\documentclass[a4paper,twocolumn]{article}
\usepackage{graphicx}
\usepackage{amssymb}
\usepackage{url}

\begin{document}
\title{The strange case of Dr. Petit and Mr. Dulong}
\author{Roberto Piazza \\\small Dipartimento di Chimica, Materiali ed Ingegneria Chimica  \\ \small Politecnico di Milano, Piazza Leonardo da Vinci, 32 - 20133 Milano}
\date{}
\maketitle
\begin{abstract}
The Dulong-Petit limiting law for the specific heats of solids, one of the first general results in thermodynamics, has provided Mendeleev with a powerful tool for devising the periodic table and gave an important support to Boltzmann's statistical mechanics. Even its failure at low temperature, accounted for by Einstein, paved the way to the the quantum mechanical theory of solids. These impressive consequences are even more surprising if we bear in mind that, when this law was announced, thermal phenomena were still explained using Lavoisier's concept of caloric and Dalton's atomic theory was in its infancy. Recently, however, bitter criticisms  charging Dulong and Petit of `data fabrication' and fraud, have been raised. This work is an attempt to restore a more balanced view of the work performed by these two great scientists and to   give them back the place they deserve in the framework of the development of modern science.
\end{abstract}
\section{Introduction}
\label{s.intro}
On 19 April 1819 Pierre--Louis Dulong appeared before the \emph{Acad\'{e}mie des Sciences} in Paris to read a paper, jointly prepared with Alexis-Th\'{e}r\`{e}se Petit, which was going to stand as a fundamental step in statistical physics~\cite{PD1819}. What Dulong and Petit  had found was, in their own words, that ``the atoms of all simple bodies have exactly the same capacity for heat"\footnote{``Les atomes de tous les corps simples ont exactement la m\^{e}me capacit\'{e} pour la chaleur".} or, more specifically, that the product of the specific heat of 13 chemical elements times their atomic mass (hence the molar specific heat) was approximately constant.    As a matter of fact, the justification of what is now known as the Dulong--Petit (DP) limiting law for the (vibrational) specific heat was one of Boltzmann's great achievements. On the other hand, evidence of the crushing failure of the DP law at low temperatures, besides giving support to Nernst's Third Law of thermodynamics, motivated Einstein to introduce quantum concepts in condensed matter physics and  Debye to develop a consistent vibrational theory of the heat capacity of solids~\cite{Pais1982}. Even today, the study of the anomalous behavior of the specific heat close to a quantum critical point has granted the DP law a ``second wind"~\cite{Amusia2015}.

The accomplishment of Dulong and Petit is even more remarkable when framed within the historical and geographical context in which it was obtained. At that time thermodynamics, still in its infancy, was based in France on the caloric theory of Lavoisier and Laplace, Dalton had introduced his atomic theory less than two decades earlier, and Jacob Berzelius had published a first table of atomic weights only six years before~\cite{Fox1971,Brush1976}. Besides, quantitative values of the ``capacity for heat'' of bodies, a concept that had slowly developed in the late XVIII century from the discovery of latent heat by the (French--born) Scottish chemist Joseph Black, were still rather inaccurate. Indeed, the advancements both in the experimental methods and in data interpretation that, as we shall see, paved the way to the final discovery by these two young scientists (Dulong was 34, and Petit just 27), are truly impressive. And yet\ldots

Yet, in the data presented by Petit and Dulong to support their revolutionary finding there is something odd, something that does not line up completely. Elusive concerns in contemporary accounts and even in authoritative studies of their life and work~\cite{Lemay1948,vanSpronsen1967,Fox1968} recently progressed to the stage of questioning the scientific integrity of the two scientists, even of blaming them of fraud and ``data fabrication''~\cite{Giunta2002}. To be true, these harsh remarks seem to have been confined to a rather restricted audience\footnote{So far, I have not found any colleagues aware of this diatribe, with the exception of Albert Philipse, a valuable Dutch chemist at the van `t Hoff Laboratory in Utrecht, with whom I share the curiosity for the history of science.}. So, I regarded as useful to bring it to the attention of physicists, perhaps from the slightly different perspective that  I may have of the development of scientific ideas.

\begin{figure}[h!]
\includegraphics[width=\columnwidth]{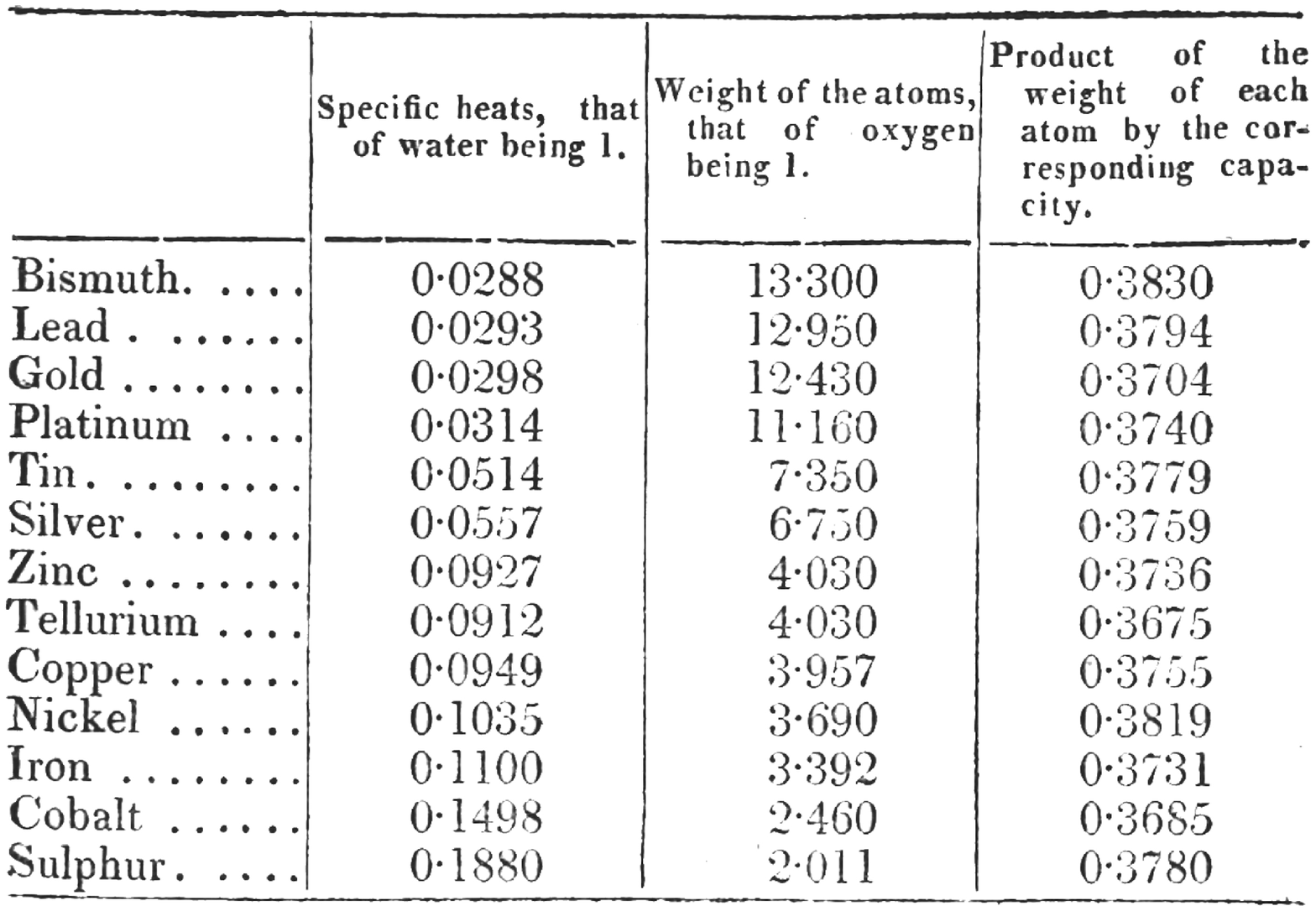}
\caption{\small Data used by Petit and Dulong to derive their law (for convenience of the English readers, I plot the version published in the contemporary translation of Ref~\cite{PD1819} that appeared in the Annals of Philosophy~\cite{PD1819b}). }
\label{f1}       			
\end{figure}

Before venturing into this enterprise, rather demanding for a physicist on the job who does not pretend at all to be a professional in historical issues, let me however tell you how I became aware of this strange story. While teaching statistical physics in my university I have always found extremely helpful to complement my technical presentation of a subject with some historical remarks that students  generally appreciate. The very early stage at which the DP law was obtained has always intrigued me, hence I decided, while writing the English version of my notes~\cite{Piazza2017}, to investigate a bit more the matter by reading the original paper by Petit and Dulong. As an experimentalist, the first thing I did was of course checking the data that they presented in the table reproduced in Fig.~(\ref{f1}), comparing at the same time their values for atomic weights and molar specific heats (at constant pressure) with modern ones.  As an appetizer, my readers may want to to repeat the same exercise. Just to give you a couple of hints, you should find out that:
\begin{enumerate}
\item Even the most trivial check, namely that Column 1 $\times$ Column 2 = Column 3 fails in one case;
\item  Much worse, even if the value of the atomic weight of at least two of the elements is largely wrong, the corresponding molar specific heats in Column 3 are quite close to modern values. Which means of course that there are compensating errors in the \emph{experimental} values of Column 1.
\end{enumerate}
There are also some additional puzzling values that, with a deeper investigation, you may have discover and that we shall discuss later. In any case, if you have done the exercise, you may understand why it did not take me long before falling in a state of deep consternation, concluding that either a) there was something fishy, or b) my neurons have already sublimated more than I am aware of. Although the second possibility could not definitely be  excluded, I went into a rather frustrating bibliographical search that lead me to discover just a single paper written in 2002 by Carmen Giunta, a professor in chemistry at Le Moyne College in Syracuse, with the rather severe and accusatory title ``Dulong and Petit: a case of data fabrication?"~\cite{Giunta2002}. Although, as you will see, I do not share Giunta's bitter criticism, I think that this work has not attracted the attention it deserves\footnote{Giunta's paper was actually presented at the 221st National ACS Meeting with the (not much lighter) title ``Dulong and Petit: a Case of Scientific Misconduct?".}.  To avoid jumping to rushed conclusions, I had to embark upon a rather long investigation that arguably provided me with a more balanced perspective of this story, and which I hope you may wish to follow. Let us first introduce the two main characters we shall deal with.

\section{A hapless physician and a boy wonder}
\label{s.bio}
Life was not kind to  Dulong, at least at the beginning. Born in Rouen on February 13, 1785\footnote{Most biographical accounts, and Wikipedia too, state that Dulong was born on February \emph{13}, but this is a longstanding historical mistake. See the results of the detailed investigation performed in 1843 by a commission of the Royal Academy of Rouen~\cite{Deville1844}, where his original \emph{act de bact\^{e}me} is presented.}, orphaned at the age of four, he was raised by his aunt and godmother Mme Faurax in Auxerre, where she took care of his education ``with all the tenderness of a mother"~\cite{Jamin1855}. Mostly by dint of his own efforts, he prepared himself for the \'{E}cole Polytechnique at Paris and matriculated at sixteen, the minimum entrance age. Yet, his studies at the \'{E}cole were plagued by sick leaves that prevented his admission to the artillery corps, so that he was eventually forced to abandon the \'{E}cole without completing the course. Dulong's impaired physical condition turned his attention to medicine, which in those days did not require lengthy or deep studies\footnote{Thus Dulong eventually became a medical doctor, but he never finished a doctorate.}. Yet, this did not turn out to be a good choice, at least financially. He started practicing medicine in one of the poorest neighborhoods of Paris $12^\mathrm{th}$ arrondissement, where, according to Arago~\cite{Arago1855},
\begin{quotation}
\noindent \emph{The clientele was increasing visibly, but his fortune diminished with the same rapidity, for Dulong never saw an unfortunate man without succouring him; because he had felt obliged to have an account open at the pharmacist, at the benefit of the patients who, without this, could not have made use of his prescriptions.}\footnote{``La client\`{e}le s'augmentait \`{a} vue d'oeil, mais la fortune diminuait avec la m\^{e}me rapidit\'{e}, car Dulong ne vit jamais un malheureux sans le secourir; car il s'\'{e}tait cru oblig\'{e} d'avoir un compte ouvert chez le pharmacien, au profit des malades qui, sans cela, n'auraient pas pu faire usage de ses prescriptions".}
\end{quotation}
Obviously Dulong could not endure this state of affairs for a long time and after some years he was forced to give up with his career as a physician. Luckily, the passion for chemistry that he had steadfastly cultivated during his medical studies draw the attention of Louis-Jacques Th\'{e}nard, who took him as \emph{r\'{e}p\'{e}titeur} of his course in \'{E}cole Polytechnique. The real turning point in Dulong's life took place when Th\'{e}nard introduced him to Claude--Louis Berthollet, who recognized at once Dulong's talent and invited him to become a member of the \emph{Societ\'{e} de l'Arcueil}, a kind of ``country club" of French scientists that he had created together with Laplace. For our purposes, it is worth noticing that the Society of Arcueil, which  besides Laplace included among its members great physicists such as Biot, Gay--Lussac, Arago and Poisson, was at that time a stronghold of Lavoisier's caloric theory, where the ``vibrational" (kinetic) theory of heat was strongly opposed and Dalton's atomic theory was still far from being generally accepted~\cite{Fox1968,Fox1971}: The discovery by Dulong and Petit was going to change this mood drastically.  In Arcueil Dulong performed his first important chemical study in which he extended Berthollet's studies of salt decomposition to show that also insoluble salts are capable of exchanging constituents with soluble electrolytes~\cite{Dulong1812}, a study that fostered the development of the law of mass action~\cite{Lemay1948,Holmes1962}. Dulong began to gain a reputation as a brilliant and extremely careful experimentalist swiftly. However, life was not tired to ambush him yet. In October 1811 the synthesis of $\mathrm{NCl_3}$, which turned out to be one the most violently explosive substance ever discovered, costed him two fingers, so that he managed to resume his work only after several months\footnote{Nitrogen trichloride has been a bane for other scientists too. Intrigued by Dulong's discovery, Humphrey Davy repeated the synthesis: this led to another explosion the that lodged a piece of glass in his cornea. In a pitiful way, this was beneficial to the development of science, since it is because of his temporary blindness that Davy was forced to hire a young assistant: Michael Faraday. But Faraday was not exempt from the curse of $\mathrm{NCl_3}$ either: An explosion that took place while he was holding a test-tube containing grains of nitrogen trichloride tore away his nails and burnt his fingers so much that he was unable to use them like before for a long time.}. Anyway, his academic career was settled: in 1820 he became professor of chemistry at Sorbonne, and then professor of physics at \'{E}cole Polytechnique replacing exactly Petit after his untimely death (see below). He concluded his life as Director of Studies of the \'{E}cole, respected if not exactly popular as a professor\footnote{His lectures do not seem to have been exactly exciting, for he despised display and regarded any extra word as a word wasted~\cite{Lemay1948}.}, dying of cancer on July 19, 1838.

Biographical accounts of the life of  Petit are much poorer, and substantially based on the funeral speech delivered by Biot~\cite{Biot1821}. For sure we can say that, compared to Dulong's life, it was almost all the way around. Born in Vesoul (not far from Besan\c{c}on) on October 2, 1791, he soon proved to be a child prodigy by completing all entrance requirements of the \'{E}cole Polytechnique before turning 11. Even for a boy wonder this was too early to be admitted to the most prestigious school in France, so he was placed in a preparatory school to fill in the time with math and literature before reaching the minimal admission age of 16. Petit finished his two--year course at the \'{E}cole with unheard-of distinction,  graduating \emph{hors de ligne} (namely, he outranked completely all of his classmates) in 1809, while it took him just another two years to get his doctorate. Leafing through his thesis, which concerns and  extends Laplace's theory of capillarity~\cite{Petit1813}, one cannot avoid being deeply impressed by the mathematical skill, the clarity, the elegance of presentation displayed by a 20 years old lad\footnote{Among the other results, Petit derives an expression for the contact angle of a liquid on the wall of a capillary in terms of the ratio between the adhesion force between the wall and the liquid and the internal cohesion of the liquid itself, providing also the condition for complete wetting. Curiously, he was apparently not aware of the basic equation already derived by Young in 1805.}. Such a remarkable talent allowed Petit to be nominated adjoint professor at the \'{E}cole Polytechnique when he was just 23 and and was advanced to the full professorship of physics (\emph{professor titulaire}) the following year (1815). A few months earlier he had become brother in law of Arago (their spouses were sisters), with whom he developed a close and friendly relationship\footnote{Arago and Petit also shared a common view about the way mathematics should be taught to students in engineering. In fact, they jointly filed a complaint about Cauchy, expressing concerns towards his insistence on teaching subjects ``which mainly had to do with series and which the students would never have occasion to use in the services'', at the detriment of applied calculus. Because of this criticism, the great mathematician eventually received a severe reprimand from the Director of the \'{E}cole.~\cite{Belhoste1991}.} and wrote his first important paper about the refractive power of substances in different states of aggregation~\cite{Arago1816}. In a few years, Petit was already regarded as one of the most brilliant promises of French science, but unhappily his potential was never fully realized. The sudden death of his wife in 1817, just six months after their marriage, drove Petit to a deep state of depression in which he showed exhibited alarming symptoms of premature senescence\footnote{About the state of health of Petit in his last days, Dulong wrote to Berzelius: ``\emph{He never realized his condition. However, his last days were painful for me. His sickness had changed his character, he acquired an aversion for everyone else around him; I was the only one who retained his confidence, and he demanded that I stay near him to talk with him about his health, during whatever spare time my duties allowed me. The grief that this sight brought me, together with the fatigue from my own work, have profoundly affected my health. Perhaps I am destined to follow him soon and by the same road}"~\cite{Lemay1948}.}. In a short time, he was no longer able to teach. In June 1820, when he was not yet thirty, Petit died from tuberculosis, the same disease that carried away his beloved spouse~\cite{Biot1821}. It is at least comforting to observe that the great result ha obtained with Dulong brought him undying fame.

It may useful adding to this biographical facts some ideas about the very different characters and way of working of Dulong and Petit, which I found in a more informal account written in 1855 by Jules Jamin~\cite{Jamin1855}, at that time professor of physics in the \'{E}cole\footnote{To my knowledge this paper, which is also a beautiful example of popular science writing, went so far unnoticed}. According to Jamin:
\begin{quotation}
\noindent \emph{Petit had a lively intelligence, an elegant and easy speech, he seduced with an amiable look, got easily attached, and surrendered himself to his tendencies rather than governing them. He was credited with an instinctive scientific intuition, a power of premature invention, certain presages of an assured future that everyone foresaw and even desired, so great was the benevolence which he inspired.
Dulong was the opposite: His language was thoughtful, his attitude serious and his appearance cold\emph{[\ldots]} He worked slowly but with certainty, with a continuity and a power of will that nothing stopped, I should say with a courage that no danger could push back. In the absence of that vivacity of the mind which invents easily, but likes to rest, he had the sense of scientific exactness, the gusto for precision experiments, the talent of combining them, the patience of completing them, and the art, unknown before him, to carry them to the limits of accuracy\emph{[\ldots]}
Petit had more mathematical tendency, Dulong was more experimental; the first carried in the work more brilliant easiness, the second more continuity; One represented imagination, the other reason, which moderates and contains it.}\footnote{``Petit avait l'intelligence vive, la parole \'{e}l\'{e}gante et facile, il s\'{e}duisait par des dehors aimables, il s'attachait ais\'{e}ment, et s'abandonnait \`{a} ses tendances plut\^{o}t qu'il ne les gouvernait; on lui reconnaissait une facilit\'{e} d'intuition scientifique en quelque sorte instinctive, une puissance d'invention pr\'{e}matur\'{e}e, pr\'{e}sages certains d'un avenir assur\'{e} que chacun pr\'{e}voyait et m\^{e}me d\'{e}sirait, tant \'{e}tait grande la bienveillance qu'il avait su inspirer.
Dulong \'{e}tait tout l'oppos\'{e}; son langage \'{e}tait r\'{e}fl\'{e}chi, son attitude grave et son apparence froide[\ldots] Il travaillait lentement, mais avec s\^{u}ret\'{e}, avec une continuit\'{e} et une puissance de volont\'{e} que rien n'arrêtait, je devrais dire avec un courage qu'aucun danger ne faisait reculer. \`{A} d\'{e}faut de cette vivacit\'{e} de l'esprit qui invente ais\'{e}ment, mais qui aime \`{a} se reposer, il avait le sentiment de l'exactitude scientifique, le go\^{u}t des exp\'{e}riences de pr\'{e}cision, le talent de les combiner, la patience de les achever, et l'art, inconnu jusqu'\`{a} lui, de les porter jusqu'\`{a} la limite possible de l'exactitude[\ldots]
Tels sont les traits principaux de ces deux hommes c\'{e}l\`{e}bres. Petit avait plus de tendance math\'{e}matique, Dulong se montrait plus exp\'{e}rimentateur; le premier portait dans le travail plus de facilit\'{e} brillante, le second plus de continuit\'{e}; celui-l\`{a} repr\'{e}sentait l'imagination, celui-ci la raison, qui la mod\`{e}re et la contient."}
\end{quotation}
This lively portrait of the two scientists, and in particular Jamin's closing sentence, may give us some clues about the curious story of their great discovery.

\section{Prelude to the discovery and a further DP law}
\label{s.first}
It is not clear how Dulong and Petit came into contact. Very likely, the \emph{trait d'union} between them has been Arago, member of Arcueil (to which Petit has never been associated) and professor of analytic geometry\footnote{Arago succeeded Gaspar Monge, the father of differential geometry who was very influential for the mathematical education of Petit.} at the \'{E}cole (where Dulong was just a modest \emph{examinateur}). As we shall see, Arago may have played a subtle and crucial role in the announcement of the DP law too. We know however of a precise event that may have stirred up their common interest towards problems related to heat transfer, namely the announcement of the subjects for prizes to be awarded by the First Class of the French Institute in 1816--1817 made on 9 January 1815~\cite{Fox1968}. Applicants were supposed to find: i) the expansion of mercury in a thermometer between $0^\circ$C and $200^\circ$C; ii) the law of cooling of a body in a vacuum; iii) the laws of cooling in air, hydrogen and ``carbonic acid" (actually, $\mathrm{CO_2}$) for several values of temperature gas density. Five month after, Dulong and Petit had already presented at the \emph{Institut de France} a joint memory that partly addressed the first of these question, but they pointed out that determining the law of cooling required a better definition of a rational temperature scale, a problem that called for further studies\footnote{It is interesting to note that, although the subject of the competition was not related to the problem of heat capacity, Dulong and Petit already pointed out the crucial importance of checking for any temperature variation of the specific heat when developing thermometers based on the thermal expansion of a substance.}. Surely they planned to carry on this investigation, but the political and military turmoil that followed\footnote{Recall that was the year of  \emph{Les Cent-Jours} of Napoleon and of the following rebellion against restoration that ended with the Treaty of Paris.} forced them to give up the competition and to publish the following year a rather short paper that was nothing more than their original memory~\cite{DP1816}. Luckily for them, no submission was judged to be worthy enough to award the prize, so that the same subject, with little but not irrelevant adjustments\footnote{Determination of mercury expansion should have been performed down to $-20^\circ$C and compared with that of an air thermometer. For the announcement, see J. Chim. Phys., \textbf{4}, 302-303.}, was proposed for the next year.

This time, Dulong and Petit produced a masterpiece~\cite{DP1818}. In a paper published in three parts (adding up to a total of 113 pages), which was to be acclaimed for a long time as a model of experimental method, they carefully and extensively address all the questions presented above with unprecedented rigour, remarkable clarity of presentation, extremely careful data analysis. A detailed and authoritative account of this monumental work has been given by Robert Fox~\cite{Fox1968}. Here I shall only expatiate on two aspects that highlight the pivotal role played by the synergy of Dulong's experimental ingenuity and accuracy with Petit physical intuition and mathematical rigour.

While introducing their investigation of mercury expansion, Dulong and Petit immediately highlight a delicate aspect. To determine the volume expansion rate of a fluid, just measuring the elongation of the latter enclosed in a glass tube with temperature like in a standard thermometer is far from being an  accurate method. Indeed, the tube expands too, hence what is measured is just a \emph{relative} thermal expansion, which also depends on the material the tube is made of. How can one get rid of this annoying problem and get an \emph{absolute} expansion rate? The brilliant method devised by Dulong and Petit was absolutely new\footnote{Actually, Dulong and Petit, who are very careful in citing previous works, state that this method was originally suggested by Boyle, and that several other scientists had \emph{thought} of using it. Yet, they point out that its practical application is far from being  easy, in particular for large temperature differences. As a matter of fact, no one used it before them.}. In their words (Ref.~\cite{DP1818}, pag. 125):
\begin{quotation}
\noindent \emph{\emph{[The method]} is  based on this incontestable principle of hydrostatic, that when two liquid columns communicate between them by a lateral tube, the vertical heights of these columns are precisely in inverse ratio to their densities. Therefore, if one could exactly measure the heights of two columns of mercury contained in the branches of an inverted glass siphon, by keeping one in melting ice, for example, while the other is brought to a known temperature, one would easily deduce the sought dilation. In fact, if $h$ and $h'$ are the vertical heights of two columns giving the same pressure at temperatures $t$ and $t'$, we must have, by calling $d$ and $d'$ the corresponding densities\mbox{, $hd = h'd'$}.}
\end{quotation}
In simple words, they devised an instrument working on the principle of the barometer that is totally independent from the material, size or uniformity of the columns and moreover yields a \emph{differential} measurement of the density: I am sure that any experimentalists will immediately appreciate this last major advantage. The following 8-page long description of the set up they built and of experimental protocol they applied is by itself a piece of bravura. Of particular interest is the accuracy with which they measured the column heights using a cathetometer equipped with a vernier that allowed appreciating displacements as small as  $0.02\,\mathrm{mm}$, probably built for them by the great instrument--maker Henri--Prudence Gambey (see Ref.~\cite{Arago1855}, pag. 604).
With this instrument they measured the absolute volume expansivity $\alpha_V$ of mercury in a wide temperature range ($0-300^\circ$C) with a much higher accuracy than in previous studies (their data are actually within a fraction of a percent with modern ones\footnote{It is worth noticing that, at a time when statistics was still in its infancy (Gauss introduced the concept of ``mean error", the forerunner of standard deviation, only a few years later), Dulong and Petit provide not only ``la moyenne d'un grand nombre de mesures", but also ``les valeurs extr\^{e}mes", namely the full range of values they obtained (which was of the order of $10^{-3}$ of the averages).} ). This also allow them to show the error that would be made by assuming a $T$-independent expansivity: for instance, a mercury thermometer that agrees with an air thermometer at $0^\circ$C would be in differ from the latter $14^\circ$C by more than at $300^\circ$C. But this is not the whole story. Dulong and Petit went on by showing that they could simply obtain the value of $\alpha_V$  for a \emph{solid} material by contrasting the absolute expansivity of mercury with its apparent one in a tube made of that material. With this brilliant trick, Dulong and Petit obtained the thermal expansivity of glass, iron, copper, and platinum.

While one can almost hear the voice of Dulong describing the investigation of thermal expansivity, the deep physical intuition and the great mathematical skill of Petit stand out when we consider the second part of the paper, dedicated to the determination of the law of cooling. This subject was arguably the main goal of the competition, motivated by the controversies concerning Fourier's model of heat conduction in solids for which he had been awarded the 1811 prize of the Academy\footnote{It is probably because of the severe criticisms raised by the committee that judged this work, which also included Laplace, Lagrange and Legendre, that Fourier's law appeared in print only several years later~\cite{Fox1968}.}. What was actually asked to the competitors was to investigate how heat is transferred in vacuum and in fluids, with the main aim of proving or disproving the “Law of Cooling” enunciated by Newton, which states that the time it takes for a sample to cool is proportional to the temperature difference with the surrounding environment.

Actually, Newton's meditations and experiments
about cooling cannot be exactly regarded as ``crystal clear". In his original sparse writings about the subject, the great scientist does not distinguish between the different modes of heat transfer. In fact, if a sample is left to cool freely, generating spontaneous (``natural" or ``free") convective currents in the surrounding air, Newton's law does not hold as a rule, while it does in the presence of \emph{forced} convection, for instance if we blow fresh air around the sample with an hairdryer\footnote{Technically, this is because in forced convection the Nusselt number, which is proportional to the ratio $\dot{Q}/\Delta T$ between the heat transfer rate and the temperature difference, must be a function of the Reynolds and Pecl\'{e}t numbers alone, which are both $T$-independent. Conversely in natural convection, where no characteristic velocity scale (and therefore a meaningful Reynolds number) can be defined, the heat transfer coefficient $h= \dot{Q}/A\Delta T$, where $A$ is the heat transfer surface, is in general a function of $\Delta T$, unless the latter is small. For instance, in the simple case of free convection from a vertical plate, $\dot{Q} = (\Delta T)^\alpha$, where $\alpha$ increases from 5/4 for laminar flow to  4/3 for fully turbulent flow~\cite{Tritton1988}.}   Newton was probably aware of the problem when, in his \emph{Scala graduum caloris}~\cite{Newton1701}, he wrote that the piece of iron he was studying was laid not in calm air but in a wind that blew uniformly on it\footnote{``\emph{Locavi autem ferrum, non in aere tranquillo sed in vento uniformiter spirante ut aer a ferro calefactus semper abriperetur a vento \& aer frigidus in locum ejus uniformi cum motu succederet}".}. Yet, this sentence was probably too cryptic to be adequately appreciated. In fact, most of the claims of failure of the Newton's law made in the $18^{th}$ century can hardly be trusted, because of the poor definition of the experimental conditions, adding up with a substantial degree of confusion about the concept of convection as opposed to conduction.

Newton himself, however, had already noticed in his \emph{Opticks} (Query 18) that a cooled thermometer heats up even if enclosed in a transparent vessel wherefrom air has been pumped out. In his own words:
\begin{quotation}
\noindent \emph{Is not the Heat of the warm Room convey'd through the Vacuum by the Vibrations of a much subtiler medium than Air, which after the Air was drawn out remained in the Vacuum? And is not this Medium the same with that Medium by which Light is refracted and reflected, and by whose Vibrations Light communicates Heat to Bodies, and is put into Fits of easy Reflexion and easy Transmission?}
\end{quotation}
In other words, with a strike of genius Newton had fully realized that heat can also be transported by something akin to light, namely, he had discovered what later was called ``radiant heat''.

\begin{figure}
\includegraphics[width=\columnwidth]{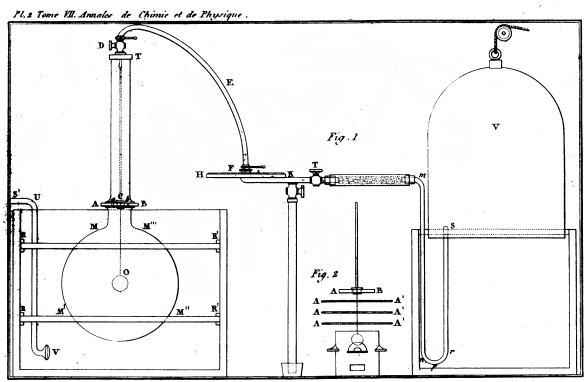}
\caption{\small Apparatus used by Dulong and Petit to investigate cooling in vacuum and in air. For a detailed (almost an understatement) description of the setup, see Ref.~\cite{DP1818} (source: Hathi Trust Digital Library, from the original conserved in the library of the University of Virginia.)}
\label{f2}       			
\end{figure}

Dulong and Petit begin their discussion by clearly stating that the process of cooling in air takes place via two distinct mechanisms, radiation and heat transfer by the fluid, which may obey different laws and must therefore be separately investigated. They also attentively point out some general experimental requirements, in particular that during cooling the sample must be thermally homogeneous, a condition that is more easily satisfied  for small samples of \emph{liquids}, so to take advantage of convection for leveling out temperature differences. The simplest choice was studying the cooling rate in vacuum of mercury contained in the glass bulb of a thermometer. In order to obtain a general picture, however, they subsequently scrutinize the effects of a) the sample volume, b) the nature of the investigated liquid (by comparing mercury to water, ethanol, and sulfuric acid, and  c) the container shape (spherical vs. cylindrical) and material (a glass vs. an iron sphere). Thanks to this thorough investigation, they manage to reach a key conclusion, namely that ``the law of the cooling of a liquid mass, variable with the state of the surface which serves as its envelope, is nevertheless independent the nature of this liquid, the shape and size of the vase that contains it''. The original drawing of the setup they used is shown in Fig.~(\ref{f2}).

What they still had to do was finding that law, and here is the genius of Petit that stands out. There was already evidence, in particular thanks to careful experiments performed by John Leslie, that Newton's law of cooling does not hold when the temperature difference $\Delta T$ between the sample and the environment is large\footnote{Leslie gave a key contribution to the study of radiant heat by observing and quantifying the different emissivity of surfaces (for instance black vs. metallic) using the ``photometer'' he invented (later called the ``Leslie cube"). Curiously, although he held that light and heat were only different states of the same substance, he claimed that heat could never be transmitted through a completely empty space, stating that ``\emph{Were it possible to procure an absolute vacuum, a body thus insulated would indisputably retain for ever the same temperature}" (see Ref.~\cite{Leslie1804}, pag. 142). Radiant heat also played an important role in the controversy between the caloric and the vibrational theories of heat (for extensive accounts, see~\cite{Fox1971,Brush1976}).}, and their experiments fully confirmed that the cooling rate increases with $\Delta T$ more than linearly. They clearly realize that they could have easily fitted the data by adding terms of higher order in $\Delta T$, but they argue that formulas of this kind, useful with no doubt when one needs to calculate intermediate values within the interpolated range, almost always become inaccurate outside the limits between which they have been determined, and are never able to make known the \emph{laws} of the phenomenon under study. In fact, as they state, their study had led them to develop a comprehensive ``theory of irradiation", something that, they believe, no physicist had done before.  This is truly a qualitative leap, which to me amounts to recognize the difference between ``taking measurements" and performing an \emph{experiment}.

To reach their goal, Dulong and Petit first use the notion, qualitatively introduced by Prevost in 1791~\cite{Prevost1791}, that equilibrium in radiation exchange comes from a \emph{balance} between emission and absorption\footnote{``\emph{Dans la th\'{e}orie adopt\'{e}e des \'{e}changes de chaleur, le refroidissement d'un corps dans le vide n'est que l'exc\`{e}s de son rayonnement sur celui des corps environnans}", see~\cite{DP1818}, pag. 148.}. Hence, accounting both for the radiation emitted by the sample at temperature $T+\Delta T$  and that received from the environment at temperature $T$, they write the cooling rate as $v = \mathrm{d}T/\mathrm{d}t = F(T+\Delta T)-F(T)$, where the unknown function $F(T)$ is in fact the ``radiation law" they were looking for. Of course, one recovers Newton's law\footnote{Which Dulong and Petit actually call ``la loi de Richmann,", from the extensive experiments by Georg Wilhelm Richmann for water cooling in air, which fully supported the law~\cite{Besson2012}.} only if $F(T)$ is linear in $T$. On the contrary, however,
according to Dulong and Petit their experimental results could be summarized by stating that ``the cooling rate of a thermometer in the vacuum increases in geometric progression when the temperature of the enclosure increases in arithmetic progression" and that ``the ratio of this geometric progression is the same whatever the excess of temperature considered''. With a simple but not trivial calculation, Petit shows (it must have been him, necessarily!) that this implies \mbox{$F(T) \propto a^T +c$}, where $a$ is a \emph{universal} constant that their data indicated to be $a=1.0077$, and $c$ another constant that can be neglected by appropriately choosing the zero of the temperature scale~\footnote{Dulong and Petit were aware that this exponential form for $F(t)$ implies that the ``absolute zero" temperature (still hypotetical at that time) must be $-\infty$, but observe that this does not necessarily means that a body contains an \emph{infinite} amount of heat. To avoid this, indeed, is sufficient that the integral over $T$ of the specific heat (which they already know to decrease with $T$) is finite.}. Consequently, $v \propto a^T(a^{\Delta T} -1)$. In the last part of their paper Dulong and Petit comply with the final requirement of the competition by performing extensive cooling measurements in the presence of several gases (air, hydrogen, carbon dioxide, and ethylene) concluding that in the pressure range $p\in [45-720]$ Torr the gas yield an additional contribution to the cooling rate proportional to $\Delta T^b p^{\,c}$, where $c$ depended on the investigated gas (but, with the exception of hydrogen, was of the order of $c\simeq 0.5$), while $b$ had a \emph{universal} value $b\simeq 1.23$ (hence very close to the value expected for free convection!).

\begin{figure}
\includegraphics[width=\columnwidth]{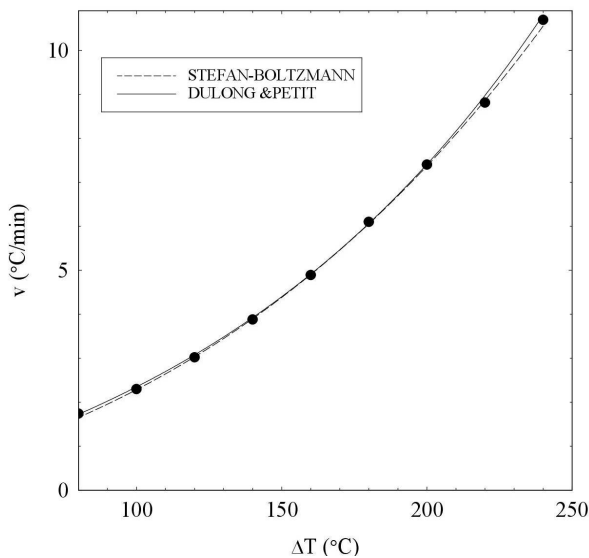}
\caption{\small Data obtained by Dulong and Petit for the cooling rate $v(\Delta T)$ of a glass bulb filled with mercury (environment at $T_0 = 0^\circ$C), with fits to the Dulong--Petits and the Stefan--Boltzmann laws.}
\label{f3}       			
\end{figure}

Curiously, till the end of the XIX century it was \emph{this} one, and not the one about specific heats, that was considered among physicists as ``the" Dulong--Petit law. As we presently know, however, this ``further" DP law is wrong, for the true expression that governs the temperature dependence of emission and absorption of radiation is given by the Stefan--Boltzmann law, which in modern terms states that the power radiated per unit surface area from a body is given by $P = \epsilon \sigma T^4$, where $\sigma \simeq 5.67\times 10^{-8}\,\mathrm{W\,m^{-2}K^{-4}}$ is the Stefan--Boltzmann constant and $\epsilon$ is the surface emissivity coefficient ($\epsilon =1$ for a black body). This implies that $F(T)$ does not increase exponentially with temperature, but as a power--law. It took however a long time before discrepancies from (this) DP law could be found (see for instance Ref.~\cite{Brush1976,Dougal1979}.). Here I wish only to recall that Stefan started the investigation that led him to propose in 1879 the $T^4-$law by arguing that the vacuum level that Dulong and Petit could reach (about $2-3$ mm of Hg) was not sufficient to ensure that the residual gas did not contribute to heat transport. Stefan indeed agreed  that convection becomes negligible at these pressures, but heat conduction, according to Maxwell kinetic theory, does \emph{not} depend on $p$ over a large range, and should be taken into account. Nevertheless, Stefan himself had to admit that the law he was proposing did not give a much better fit to the DP data (which evidently were still regarded as a reference 60 years later), and only an order of magnitude comparison with the data obtained by Tyndall for a platinum wire heated up to $1200^\circ$C allowed him to state that, on a wider $T$-range, his law was much better obeyed~\cite{Brush1976}. Just to highlight how hard was to disprove the DP law of cooling, I have contrasted in Fig.~(\ref{f3}) one of their original fits with a fit to the same data made using the correct Stefan--Boltzmann law.

I have expatiated a lot on the 1817 paper, first because it conveys a clear, undeniable message: Dulong was an exquisite experimentalist, and Petit a first-rate theorist. But there is a second reason, which is no lesser importance for our purposes. Even if the competition announced by the French Institute had little to do with the heat capacity, Dulong and Petit, motivated by the crucial role of the latter in establishing a proper thermometric scale, \emph{did} measure specific heats using the (now) classical method of mixtures. Right at the end of the first part of their paper they first obtain the specific heat $c_p$\footnote{\label{cp}I use the symbol $c_p$ for the specific heat at constant pressure, which is of course what Dulong Petit were really measuring, that is related to $c_v$ by $c_p = c_v +\alpha^2T/(\rho\beta_T)$, where $\alpha$ is the thermal expansivity, $\beta_T$ the isothermal compressibility, and $\rho$ the mass density of the material. It is worth noticing however that they were already aware of that $c_p \ne c_v$, since this has already pointed out by Dalton. However, they correctly estimate that the amount of heat spent for volume expansion of an almost incompressible solid or liquid can be safely neglected.}   of iron within four different temperature ranges between $0^\circ$C and $350^\circ$C and then the specific heat of mercury, zinc, antimony, silver, copper, platinum, and glass for two different $T$-ranges. Dulong and Petit were indeed clearly aware that $c$, like the thermal expansivity, \emph{does} depend on $T$\footnote{``Il en est donc des capacit\'{e}s des corps solides comme de leurs dilatabilit\'{e}s; elles croissent avec les temp\'{e}ratures mesur\'{e}es sur le thermom\`{e}tre \`{a} air", Ref~\cite{DP1818}, pag. 147.}. But there is no attempt of finding any relation among these values,  no sign of the simple but great insight that came to their mind roughtly a year later, although some of the values for the specific they had obtained may have hinted at that conclusion. Why it did not happen is the first of several little puzzles we shall encounter in the next sections.

\section{Everlasting fame: the great achievement of 1819}
The first odd thing about the paper that gave our two scientists everlasting fame is the order of names. In all  previous papers, Dulong had been the first author, and it was \emph{Dulong} who read the paper at the \emph{Acad\'{e}mie des Sciences}. After all Dulong was much more experienced and recognized them Petit, not to say that he was also six years older. But the DP law should properly be called the law of \emph{Petit and Dulong}, for the first author of Ref.~\cite{PD1819} is Petit\footnote{I made a little effort to patch over this historical injustice with the short title of this paper (see the top of this page)}. This already suggests that it was the young rising star that had the brilliant intuition of checking the products $c_p\times m_a$ of the specific heats times the atomic weights. But a hypothetical backstory, advocated by Jean--Baptiste Dumas, may also suggest that the two scientists had a dissimilar opinion about the experimental evidence supporting their finding~\cite{Fox1968}. According to the eulogy of Henri--Victor Regnault he presented in 1881\cite{Dumas1883}, on 5 April 1918 (``a memorable date", according to Dumas) Petit confidentially (and eagerly) showed his brother-in-law Arago a piece of paper where he had discovered the unique similarity of the values of $c_p\times m_a$. Arago immediately grasped the importance of the discovery but, apparently, he had good reasons to suspect that Dulong might have objected to publication. Hence, to convince him, he leaks the news to his fellow members of the Acad\'{e}mie. His stratagem evidently worked out, since it did not take too for Dulong to present the joint paper to the Academy\footnote{``\emph{Une heure apr\`{e}s, l'illustre secr\'{e}taire perp\'{e}tuel convaincu que Dulong, toujours h\'{e}esitant, pourrait s'opposer \`{a} la divulgation de cette belle loi, en entretenait ses confr\`{e}res, par une indiscr\'{e}tion calculee. Huit jours plus tard, les deux collaborateurs l'\'{e}non\c{c}caient devant l'Acad\'{e}mie elle--m\^{e}me}\ldots"\cite{Dumas1883} (Note that Dumas speaks of 8 days, while they actually seem to have been 14).}.

A story told more than 60 years later by someone who in 1819 was just a student in Geneva may sound apocryphal, even more because no other account of this meeting between Petit and Arago survives. Yet, Dumas had close ties with Arago, who may have been the original source, and his interest for the work of Petit and Dulong dates back to 1826, al least\footnote{\label{Dumas}Actually, in Ref.~\cite{Dumas1883} Dumas claims to have been the one who spurred Regnault to carry on the work of Dulong and Petit. Dumas also developed an interesting method for measuring the molecular weight of volatile substances, which basically consists in placing a small quantity of the that substance into a flask of known volume, which is heated until the substance turns into a vapour that replaces the air in the flask: When the the substance has fully evaporated, the vessel is sealed, dried, and weighed. Unfortunately, as the shall briefly see, the results obtained by Dumas with this method, which has been a standard in organic chemistry for a long time, were one of the main reasons why Avogadro's fundamental insight was rejected by most chemists, first of all Berzelius (a full account of this story can be read in~\cite{Nash1957}).}. Whatever the truth, this curious anecdote would not be out of keeping with the observations of Jamin on our two scientists.

What is surely true is that the 1819 paper drastically differs from the previous works of Dulong and Petit in content, style, and length too, being only 19 pages long. Petit and Dulong begin with a very general incipit. After stating that they are persuaded that certain properties of matter would appear in simpler form and would be expressed by less complicated laws if one could relate them to the ``elements on which they are immediately dependent'', they indeed claim (clearly referring to the content of the paper) that
\begin{quotation}
\noindent \emph{The success that we have already attained makes us hope that this kind of reasoning will not only contribute to the ultimate progress of physics, but that also the atomic theory will in its turn receive from it a new degree of probability, and that it will there find sure criteria for the distinction of the truth among hypotheses that appear to be equally probable.}\footnote{This quotation and all those that follow are from the contemporary English translation, Ref.~\cite{PD1819b}.}
\end{quotation}
Such a bold statement implies that they were fully aware of the great importance of their discovery. Then, they specify that the attributes of matter they will focus on are those that ``depend on the action of heat'', and in particular the specific heat. However, a terse review of previous investigations of this subject lead them to conclude that ``\emph{The attempts hitherto made to discover some laws in the specific heats of bodies have then been entirely unsuccessful}". Dulong and Petit identify the origin of this failure in the difficulty of finding accurate methods of measurement. They admit that, among the proposed approaches, the method of mixtures ``may doubtless, when properly conducted, lead to very exact results". Yet, it suffers from a major drawback: it requires a sizeable amount of the investigate material, which prevents its application to rare substances (or to expensive ones, like gold or platinum!). Nevertheless, they claim, the experience they made allowed them to single out a method that satisfies all critical requirements: the method of cooling. This statement may sound a bit singular, since it seems to be based on Newton's assumption that in their previous work they had shown not to be an exact law. Yet, they are fully aware that it does apply for sufficiently small temperature differences. Accordingly, they write, ``all our experiments were made in an interval of temperature included between $10^\circ$  and  $5^\circ$ centigrade of excess above the ambient medium". Operating around the same temperature also allows them to get rid of errors resulting from the graduation of the thermometer.

\begin{figure}
\includegraphics[width=\columnwidth]{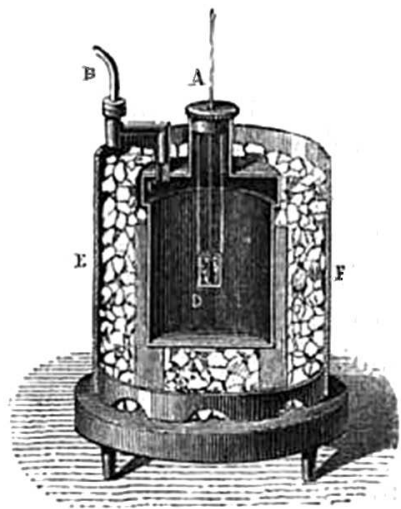}
\caption{\small Sketch of the novel apparatus discussed by Petit and Dulong in Ref.~\cite{PD1819}, as shown in Ref.~\cite{Jamin1886}, Deuxi\`{e}me Fascicule, pag. 29 (freely downloadable from The Internet Archive.)}
\label{f4}       			
\end{figure}

Although concise, the following discussion of the specific choices they made and of the apparatus they developed shows how much Petit and Dulong had learnt from their previous investigations. First, they observe that any precaution to improve temperature measurements would be ``delusive'' if the ambient temperature were not rigorously constant during the total duration of each experiment. To ensure this they plunged  the samples into a vessel, blackened in the inside and surrounded by a thick coating of melting ice. Blackening the inside of the vessel had also the advantage of slowing down the sample cooling, an important problem when working with small samples. To further reduce the cooling rate they exploit another evidence they found in their previous study, namely that ``the velocity of cooling of a body may, ceteris paribus, be considerably diminished when its surface possesses but a very weak radiating power, and is plunged in an air very much dilated." To this aim, they finely ground a tiny amount of each investigated substance, pressing then the powder into a small and thin cylindrical silver vessel with high surface reflectivity, the axis of which was occupied by the thermometer that served to
evaluate the rate cooling rate. By these precautions they managed to work with samples weighting less than 30 g even for very dense metals like platinum, still retaining cooling times of tens of minutes. Unfortunately, at variance with their previous paper (see Fig.~(\ref{f2})), they do not include any picture of their setup. However, a drawing reproducing the latter that may have been kept at the \'{E}cole can be found in Ref.~\cite{Jamin1886}  and is shown in Fig.~(\ref{f4}).

The following part of Ref.~\cite{PD1819} shows the largest departure from the style of presentation of their previous works. Indeed, there is no trace of  the minute data presentation and analysis that characterize their prize--winning publication. They actually `apologize' by stating that
\begin{quotation}
\noindent \emph{It would now be requisite to give the formula which served for the calculation of the observations; but the details into which we should be obliged to enter respecting the manner of making the different corrections depending on the method of proceeding would lead us into a discussion which we reserve for the publication of the definitive results of all the direct experiments which we have made on the subject.}
 \end{quotation}
 So, Petit and Dulong maintained that a forthcoming extended paper will have clarified any questions or misunderstandings about their findings. We know however that things went differently, due to the tragic end of Petit. Yet, we may wonder why Dulong did not keep this promise, at least to keep alive the memory of his good friend. Next comes the  climax of the paper, with the presentation and an insightful discussion of the table shown in Fig.~(\ref{f2}), in which Petit and Dulong show that they are fully aware of the terrible blow their discovery gives to the hypothesis of the caloric, and at the same time of the remarkable support it provides to the atomic theory\footnote{\label{lettre}On 15 January 1820, Dulong wrote to Berzelius ``\emph{We had already given a fatal blow to the chemical theory of warmth in the memory we read at the Institute during your stay in Paris}" and that ``\emph{Despite the objections of M.Laplace and some others, I am convinced that this \emph{[atomic]} theory is the most important concept of the century and in the next twenty years it will bring about an incalculable extension to all parts of the physical sciences}"~\cite{Soderbaum1915}}.

 It is however worth pointing out what the law of  Petit of Dulong does  \emph{not} say. First, the value of the product $c_p\times m_a$ had no explanation until Boltzmann gave it a precise physical meaning\footnote{One must say that Boltzmann did not made his best effort to `advertise' his result, which is rather buried inside a paper with the slightly misleading title ``Analytical proof of the second law of mechanical heat theory from sentences on the balance of living force"~\cite{Boltzmann1871} that does not lead itself to be easily understood (to use a euphemism). A much more readable account was written only several years later by Franz Richarz~\cite{Richarz1885}. This paper, which contains a detailed model for the transfer of heat to atom vibrations and also an interesting discussion of  anharmonicity effects, was actually regarded at the end of the XIX century as the most useful source for the theoretical interpretation of the DP law (see Ref.~\cite{Lehfeldt1894}, pag. 37).}, Second, since they knew very well from their previous work that $c$ decreases with $T$, they could not regard that value as temperature independent. Finally, they did not think that they results applied only to solids but rather to any substance, simple or compound, in any state of aggregation (as they explicitly state in the paper). In fact, after Petit's departure Dulong not only tried to extend the law to gases and compounds, but also to other physical quantities besides the specific heat like the refractive index (with Arago), with little success in both cases.

\begin{figure}[h!]
\includegraphics[width=\columnwidth]{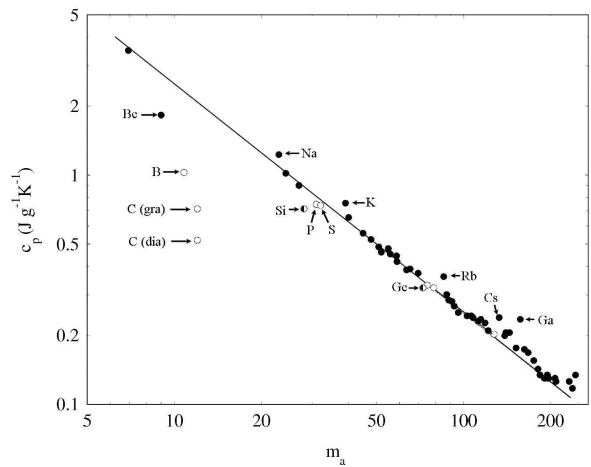}
\caption{\small Specific heat per unit mass $c_p$ of those elements that are solids at \mbox{$T=25^\circ$}C. Full and open circles respectively indicate metals and nonmetals, while the semiconductors silicon and germanium are shown by half-full dots. In this double-log plot, the law of Petit and Dulong is given by the straight line. Those elements whose heat capacity deviates appreciably from the DP law are explicitly indicated, with two allotropes of carbon, diamond and graphite, marked by C(dia) and C(gra).}
\label{f5}       			
\end{figure}

Actually, Petit and Dulong may have not even be fully convinced that what they had found was an \emph{exact} `law'. Indeed, just after the grand statement stated in the introduction, they add a caveat:
\begin{quotation}
\noindent \emph{If we recollect what has been said above respecting the kind of uncertainty which exists in fixing the specific weight of the
atoms, it will be easy to conceive that the law which we have
just established will change if we adopt for the density of the
particles, a supposition different from that which we have
chosen \emph{[\ldots]} But whatever opinion be adopted respecting this relation, it will enable us hereafter to control the results of chemical analysis; and in certain cases will give us the most exact method of arriving at the knowledge of the proportions of certain combinations.}
\end{quotation}
 \begin{table}
\caption{\small Atomic weights, normalized to the atomic weight of oxygen and rounded to the second decimal obtained by Berzelius in 1818 (B1818) and in 1826 (B1826), compared to the values used by Petit and Dulong (PD1819) and to the current values (fourth column). The last column is discussed in the next Section.}
\label{t1} 	
\footnotesize
\vspace{5pt}	
\begin{tabular}{lrrrrr}
\hline\noalign{\smallskip}  & B1818 & PD1819& B1826 & Modern & Ratio\\
\noalign{\smallskip}\hline\noalign{\smallskip}
Bi & 17.74 & 13.30 & 13.30& 13.06& 3/4\\
Pb & 25.89 & 12.95 & 25.89 & 12.95& 1/2 \\
Au & 24.86& 12.43 & 12.43 & 12.31 & 1/2 \\
Pt & 12.15& 11.16& 12.15&	12.19 & 9/10 \\
Sn &14.71&	7.35&	14.71&	7.42 & 1/2 \\
Ag &27.03	&6.75 &13.52 &6.74 & 1/4  \\
Zn &8.06&	4.03&	4.03&	4.09 & 1/2\\
Te & 8.06&	4.03&	8.02& 7.98 & 1/2 \\
Cu & 7.91&	3.96&	3.96&	3.97 & 1/2\\
Ni & 7.40&	3.69&	3.70&	3.67 & 1/2 \\
Fe & 6.78	&3.39&	3.39&	3.49 & 1/2 \\
Co &7.38&	2.46&	3.69&	3.68 & 1/3  \\
S &2.01	&2.01	&2.01	&2.00 & 1 \\
\noalign{\smallskip}\hline
\end{tabular}
\end{table}
What they suggest with these words is that their `law', even if approximate, might provide a powerful tool to solve some crucial problems in the determination of atomic weights that, at that time, was indeed far from being satisfactory. By making atomic weights the keystone of his theory and
using them as the chief criterion to set apart different atomic species, Dalton went far beyond all previous abstract and generic models, thus attracting on atoms the interest of chemists. But the few atomic or molecular weights he managed to estimate were often wrong,  also because he was using arbitrary assumptions like the `rule of greatest simplicity'\footnote{For instance, since water was known to be made of oxygen and hydrogen, Dalton assumed that its formula was OH. For a very accurate account of the development of atomic theory, see~\cite{Nash1957}.}. The true prince of atomic weight determination was Jacob Berzelius, a giant of XIX chemistry. But even Berzelius took a long time before reaching consistent and accurate values. The first and third columns of Table~\ref{t1} compare  the values of $m_a$ for the elements investigated by Petit and Dulong (normalized to the atomic weight of oxygen) obtained by Berzelius in  1818 and 1828, with the atomic weights used in Ref.~\cite{PD1819} and with the currently accepted values. It is worth noticing that Dulong and Petit were surely aware of the first set of data, although  Berzelius' work was translated in French only one year later, since Berzelius was in Paris from August 1818 to June 1819 and extensively worked with Dulong in Arcueil\footnote{Dulong and Berzelius developed a deep friendship, witnessed by the copious and warm letters they exchanged till Dulong death (see for instance footnote~\ref{lettre}). In fact, Berzelius was really fond of Dulong, and described him as ``having the most brilliant mind in the world of chemistry".}.
Table~\ref{t1} will be  useful to understand the reception and to discuss the recent criticisms of Ref.~\cite{PD1819}. Before that, however, it is useful to take a `fresh look' at the law of Petit and Dulong by plotting the currently accepted values of the specific heat at constant pressure $c_p$ versus $m_a$ for all the solid elements in the periodic table. Fig.~(\ref{f5}) shows that, apart from the a few very anomalous cases, it would be quite unfair to deny that an inverse relation $c_p\propto m_a^{-1}$ holds, at least approximately. Keeping this picture in mind, we are ready to put Dulong and Petit on trial.

\section{Petit \& Dulong on trial}
If you have done the little exercise I suggested in the introduction, checking with modern values the data in Fig.~\ref{f1}, you may have found that the most controversial results concern:
\begin{enumerate}
\item \emph{Platinum}, for which the products of the data in the first two columns gives 0.350, and not 0.374\footnote{In passing, we should not blame too much Petit and Dulong for giving the results in the last column with four decimals, when the second one has only three. After all, as I mentioned, rigorous statistics had yet to come\ldots}
\item  \emph{Cobalt} and \emph{tellurium}, whose molar heat capacities are quite close to the currently accepted values ($24.68\,\mathrm{Jmol^{-1} K^{-1}}$ to be compared with $24.81\,\mathrm{Jmol^{-1} K^{-1}}$ for Co, $24.61\,\mathrm{Jmol^{-1} K^{-1}}$ to be compared with $25.73\,\mathrm{Jmol^{-1} K^{-1}}$ for Te~\cite{CRC2007}), but whose atomic weight are underestimated by a factor of 2/3 for Co and 2 for Te (which means, of course that the experimental value are overestimated by the same factors).
\end{enumerate}
Glancing through Table~\ref{t1}, however, you will immediately notice that \emph{all} values of $m_a$ used by Petit and Dulong, except the one for sulfur, differ (in most cases quite consistently) from those just published by the great Berzelius (who, remember, may well have been hanging about their lab on these days!). Yet, look at the last column of the table: all the ratios between the values they used and those by Berzelius are almost exactly simple fractions\footnote{With the sole exception of platinum (in this case the ratio is about 0.92).}. This gives us an important clue to understand why it took so long for them to realize the evidence. Indeed, it would be silly to think that a gifted mathematician like Petit  should have taken a whole year to make $a\times b=c$ (in fact, this did \emph{not} work). What Petit brilliantly guessed is that the products of the experimental specific heats times Berzelius' atomic weights, rescaled by suitable but \emph{simple} factors, were very close. For Petit this was too nice to be fortuitous. Hence, he arguably concluded, the true values of $m_a$ \emph{must} have been the rescaled ones, which would have turned their relation into ``the most exact method of arriving at the knowledge of the proportions of certain combinations''.

This view of Petit and Dulong's results seems to have been appreciated by their contemporaries, and in particular by Berzelius himself who, seven years later, had already accepted many of the values of $m_a$ they had proposed, although he did not agree at all on the values for cobalt (rightly), silver and tellurium (wrongly). As a matter of fact, although he was not fully convinced of the generality of the law, he concludes that ``\emph{for the moment we have  to agree that a continuation of Dulong and Petit's excellent work in this subject would, however, be a vital service to science}"\footnote{My translation from he first Italian edition of Berzelius' treatise~\cite{Berzelius1839} (curiously, there was no contemporary English translation of his \emph{Opus Magnus}).}. On the other hand, Victor Regnault, carrying on their work\footnote{Regnault can actually be considered as the father of modern calorimetry. In fact, by measuring the heat capacity of about 30 elements and correcting the errors made by Dulong and Petit, he gave strong support to their hypothesis so much that he could be considered a full--fledge coauthor of the DP law.}, clearly points out that, even using the `updated' values of Berzelius, several problem persist\cite{Regnault1840}:
\begin{quotation}
\noindent \emph{Now, if we replace the atomic weights adopted by Dulong and Petit by those who are generally admitted now, we recognize that their law is far from being verified in such a satisfactory manner \emph{[\ldots]} the specific heat of bismuth is a third too weak to follow the law of atoms, the specific heat of silver and that of tellurium are twice as large; the specific heat of cobalt is too strong of about one--third; finally platinum also deviates from the theoretical number.}\footnote{``Or, si l'on remplace les poids atomiques adopt\'{e}s par Dulong et Petit, par ceux qui sont g\'{e}n\'{e}ralement admis maintenant, on reconna\^{\i}t que leur loi est loin de se v\'{e}rifier d'une mani\`{e}re aussi satisfaisante \emph{[\ldots]} la chaleur sp\'{e}cifique du bismuth est trop faible d'un tiers pour suivre la loi des atomes, la chaleur sp\'{e}cifique de l'argent et celle du tellure sont deux fois trop grandes; la chaleur sp\'{e}cifique du cobalt est trop forte environ du tiers; enfin le platine s'\'{e}carte \'{e}galement du nombre th\'{e}orique."}
\end{quotation}
As a great experimentalist, the way he settled this issue was by performing accurate measurement that rectified the values found by Petit and Dulong for Co, Te, Ag, and Bi (but also Berzelius' values of $m_a$ for the latter two elements). But he did not call into question the great value of the result they had obtained.  On the contrary, he pointed out that, since the atomic weights of the substances he investigated vary of a factor of 7 while the products $c_p\times m_a$ differ by no more than 10\%, we should be convinced that ``the law of Dulong and Petit must be adopted, if not as an absolute principle, at least as a result that approaches very much the truth''\footnote{Very interestingly, in the final part of his investigation, Regnault tries and scrutinies why the DP law is not `exact', pointing out the role of the ``\emph{chaleur latente de dilatation}" (in other words, Regnault is aware of the difference between $c_p$ and $c_v$) that, although small for solids or liquid, may produce a temperature dependence of the measured specific heat that will be different for different substances.}.

As we see, the two most interested parties, Berzelius and Regnault, knew perfectly well about the problems with Petit and Dulong's data, but did not make a big story about it. On the contrary, they were both aware of the fundamental importance this result would have, even if it might not be considered as an `absolute truth'.  Even more, it is far from them claiming any `fraud' by two scientists whom they both admired. For a long time, all respected scholars who have investigated the DP law seem to have shared this general attitude~\cite{Lemay1948,vanSpronsen1967,Fox1968}. Until, at the end of the last century, a buzz of discredit began to rise. Apparently, everything started in 1985 with a radio talk of the Australian writer Peter Macinnis, followed by a letter to Chemical \& Engineering News by Peter Schwarz~\cite{Schwarz1987}, both of which triggered the interest of Carmen Giunta~\cite{Giunta2002}. I could not listen to Macinnis' talk, broadcasted on the other side of the world when I was still a student. Yet, you can appreciate his own rather different attitude on the ABC website, where you read (\url{http://www.abc.net.au/science/slab/macinnis/story.htm}):
\begin{quotation}
\noindent \emph{Dulong and Petit concocted their results when they generated their law relating specific heat to atomic weight. Given the fraudulent data that I can demonstrate in their results, they probably faked more than half of the measurements, and fudged the rest like a second-rate physics student. But who cares? Their spurious law was more or less correct, and it allowed chemists to determine atomic weights accurately by electrolysis, ducking around problems caused by valency.}
\end{quotation}
I leave out any comments, which is left to the judgement of my readers. Similarly, I shall not waste time arguing against the offending letter by a rather obscure organic chemist. The paper by Giunta, however, deserves for sure much more attention. So, let us start by dwelling upon the three obvious inconsistencies they we already pointed out.
\begin{enumerate}
  \item \emph{Platinum.} Apparently, this is a minor problem, for most reviewers  including Giunta agree that it must just have been a misprint. Maybe, but the question is, \emph{where} is the misprint? Table~\ref{t1} suggests that it should be in the atomic weight, since this is the \emph{only} value of $m_a$ which is not a simple fraction of the Berzelius' 1818 value. Using $m_a = 12.15$, however, the product becomes 0.3818, which is not what they state (although still an acceptable value). In fact, the printer's error is in the \emph{specific heat}, for in the German translation\footnote{Curiously, in the original French publication there was another misprint, since the stated value of the atomic weight is 0.03\textbf{5}5. This misprint, corrected by Dulong and Petit themselves  in the German edition, was surely known to Regnault (see Ref.~\cite{Regnault1840}, pag.~9).} of Ref.~\cite{DP1818} Dulong and Petit found $c_p=0.0335$, which, multiplied by $m_a =11.16$, yields \emph{exactly} 0.3740. While this value (equivalent to $25.05\,\mathrm{JK^{-1}mol^{-1}}$) is about 3\% smaller than the currently accepted value, using $c_p=0.0335$ and $m_a = 12.15$ they would have obtained a molar specific heat 5\% \emph{larger} than the modern one, which is probably not a big issue. However, besides noticing that Petit and Dulong very likely ``recycled'' some of the data obtained in~\cite{DP1818}, we may wonder why they made such a strange change for the atomic weight of Pt. But here comes the real puzzle:  If they felt that the result for Pt was  suspicious,  because it obliged them to use a rather weird rescaling of Berzelius' value, why did not they simply \emph{exclude} it from their table? After all, they still had a list of 12 substances that support the law! No, they did not. Remembering how things may have gone according to Dumas, I am rather inclined to think that this might have the result of the `compromise' between Petit and Dulong who, still doubtful about some of their results and convinced that some additional work was needed, nevertheless accepted to present the paper provided that they report about \emph{all} the substances they had investigated, sweeping no dust under the carpet. Although no historical evidence will ever support my guess, this might have happened too for the other two elements we are going to examine.
  \item \emph{Tellurium and cobalt.} When compared with modern data, the Petit and Dulong results for these elements stand out as the most bewildering ones and are surely prone to rise suspicion, which eventually lead Giunta to state that ``\emph{In particular, the specific heats of cobalt and tellurium, which Dulong and Petit state they measured, appear to have been fabricated}". Before accepting this summary judgement, however, let us mull a bit more over this, taking into account the information I tried to summarized on the state of affairs in 1819. Petit and even more Dulong were surely worried about the reception of their work, where \emph{all but one} (sulfur) of the atomic weights just presented by the leading expert in the field\footnote{Looking at Table~\ref{t1}, we se that this perilous choice largely paid back, since 9 over 12 changed values of $m_a$ are pretty close to the modern ones (exception are are again Pt, Te, and Co).}.   So, in the case of Tellurium, I really do not find any reasons why they should have \emph{halved} Berzelius value for $m_a$ (which they knew) \emph{doubling} at the same time their experimental value for $c$ to make their product consistent with the others.  On the other hand, had they completely `fabricated' this result, why not taking for $m_a$ a value supported by the authority of Berzelius? Fiddling this way with data would have been, in my opinion, a clear symptom of masochism, also because tellurium was one of the substances Berzelius was more skilled with (it was because of his noticeable confidence with this substance that he managed to discover selenium in 1818)\footnote{Curiously both tellurium and selenium are \emph{really} strange elements for what concerns heat capacity. Indeed, each atoms has just two strongly bound  nearest neighbors, so the crystals resemble fibrous chain structures with weak interchain interactions. As a consequence, their low--temperature specific heats markedly deviates from the Debye $T^3$ limiting law, and vanish linearly with temperature \cite{Desorbo1953}.}. Indeed, he decidedly refused to accept the result by Petit and  Dulong result, firmly stating that ``The external properties and the specific gravity of tellurium are also similar to those of the antimony, which convince me to take their atomic weights as equal, regardless of the above mentioned experiments of Dulong and Petit"~\cite{Berzelius1839}.

      The question of cobalt is different, because in this case the Berzelius value was \emph{wrong} (of a factor of two), so any experiments must have given a conflicting value for $c_p\times m_a$. Again, if they did `fabricate' the experimental result, why not `hiding' the fraud just by using \emph{Berzelius}' value? No, they did not, they included the value for Co anyway risking to use  a \emph{different} (but wrong) value for $m_a$. Berzelius was apparently more open to questioning the value he had obtained for cobalt, but in any case in his treatise he does not seem to recommend using specific heat to find the atomic mass of tellurium, cobalt, and of the other `anomalous' elements~\cite{vanSpronsen1967}.
\end{enumerate}
As you see, the `anomalous' results for Pt, Te, and Co may have different explanations. The fact is, we are looking at these data from the advantageous point of view of the future. I really wonder if any rumors against the Petit and Dulong work had ever been raised if they did not include the only three elements they attributed to erroneous values of $m_a$. After all, hindsight is  a well known cognitive bias.

Yet, to reach his conclusions, Giunta leverages on another clue, surely more quantitative, based on comparing the statistics of the molar heat capacities obtained by Petit and Dulong with modern ones. This statistical analysis leads him to conclude that the whole paper is basically a fraud. More specifically, he claims that all Petit and Dulong had measured were just the few element discussed in~\cite{DP1818}, that everything else was fabricated, that no new experiment was performed, so that even their detailed  experimental description of the setup shown in Fig.~(\ref{f4}) (I guess for the first time after more than a century) is a fake. With what we have seen so far, however, I think we have already good reasons to refrain from making haste to such a severe sentence.  As a matter of fact, it is true that the results for iron, zinc, silver and copper are identical to those already reported in 1818. But then, why changing the value for platinum? Moreover, why not including antimony too, which they did measure in the same work? Halving Berzelius' $m_a$ as they did for many alleged `fake data', would have given $c_p\times m_a =0.408$, which is not that much higher than the other products.  Besides, with all we learned about Dulong and Petit and about the way they worked from the voices of scientists like Biot, Arago, Berzelius, Jamin, Dumas, and Regnault, Giunta's claim sounds to  me as a jarring note in a chorus of great singers.

\begin{figure}[t]
\includegraphics[width=\columnwidth]{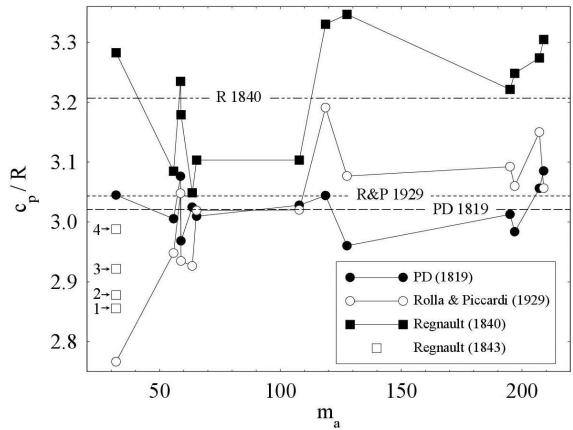}
\caption{\small Comparison of the values of the molar specific heats $c_p$ in units of the gas constant $R$ obtained by Petit and Dulong (bullets) and Regnault (squares,~\cite{Regnault1840}) with the modern ones reviewed by Rolla and Piccardi (open dots,~\cite{Rolla1929}). The dashed lines are the averages of the three data sets (excluding sulfur). Open squares are the results for sulfur obtained by Regnault in a following study (Ref.~\cite{Regnault1843}, pag. 344). Specifically, they refer to:  1) the same sample  he had studied in 1840 (crystallized from melt), measured after two years; 2) `natural' crystalline sulfur (not from melt); 3) a sample crystallized from melt after two months, and 4) just after crystallization.}
\label{f6}       			
\end{figure}

I admit, however, that rebutting (at least in part) his conclusions requires to deal with statistics. So, let us do it. Actually, the benchmark data used by Giunta are not \emph{that} modern, since he refers to the values of $c_p$ reviewed by Rolla and Piccardi in 1929~\cite{Rolla1929}, but I agree that this is apparently the most recent collection of data obtained at $0^\circ$C, which was the ambient temperature in the experiments by Petit and Dulong\footnote{More recent values, such as those reported in the CRC Handbook of Chemistry and Physics~\cite{CRC2007}, would anyway differ by, typically, 1\% or less.}. The two data sets are contrasted in Fig.~(\ref{f6}),  together with the results for the same elements obtained by Regnault in 1840~\cite{Regnault1840}, which I regard as a useful set for comparison\footnote{in the following, I shall indicate quantities relates to the modern, Petit and Dulong, and Regnault data with the subscripts 0,1, 2, respectively.}.
Before we analyze the data, however, let us consider the rather anomalous data for sulfur, the element with the lowest value of $m_a$ in the plot. In this case, the modern value is 10\% lower than the PD value and more than 18\% smaller than the value obtained by Regnault in 1840, discrepancies are far larger than those for the other elements. Sulfur is indeed a very peculiar substance, presenting two main crystalline structures, rhombic $\alpha$) and monoclinic ($\beta$), but also a spectacular polymorphism that derives from the tendency of this element to associate into homocyclic rings containing up to 20 sulfur atoms or even into long--chain ``living" polymers of indefinite length~\cite{Meyer1976,Steudel2003}. Although the $\alpha$ structure is only the low--pressure stable phase (probably made of rings containing eight sulfur atoms), fast cooling from the $\beta$ phase or from the liquid usually leads to the formation  of amorphous sulfur with a `plastic' texture, which has noticeable influence on the thermal properties of the material. In fact, it took a long time before an accurate value for the heat capacity of pure rhombic sulfur was obtained\footnote{Even the figure in Ref.~\cite{Rolla1929} differs by 3\% from the accurate value at $0^\circ$C obtained by Eastman and McGavock 8 years later~\cite{Eastman1937}.}. Notably, however, the problem was already well know to Regnault, who in a following paper briefly discussed the difficulty he met in measuring $c_p$ for this element. The open squares in Fig.~(\ref{f6}) show that a sample just crystallized from melt displays a consistently higher heat capacity, which slowly decreases in time reaching, after two years, a value very close to the one he obtained for `natural' (not melt-crystallized) sulfur\footnote{Interestingly, Regnault already suggests that this is due to `incomplete crystallization'.}. Note in particular that the value for the freshly crystallized sample is rather close to what Petit Dulong found, suggesting that the sample they used was similarly prepared.   Even the `natural' sulfur studied by Regnault  has a molar heat capacity that is 3\% higher than the modern value, which is so low with respect to the other elements in the plot because sulfur has a rather high Debye temperature, $\vartheta_D = 527\,$K~\cite{Ho1974}. In fact, sulfur contributes to about 1/3 to the fluctuations about their average of the modern data. Obviously, these fluctuations are not due to any `measurement error', but rather reflect \emph{physical} differences among the elements, mostly (but not exclusively) due to the different values of their Debye temperatures, which vary from 87\,K for Pb up to 386\,K for Co\footnote{For the modern data set, the correlation coefficient between  $c_p$ and $\vartheta_D$ (excluding again S) is about 0.57.}. What I shall call for brevity `averages' and `standard deviations' are not therefore statistical estimators of an underlying simple statistical distribution, but are just parameters quantifying the mean and the r.m.s. of \emph{intrinsic} fluctuations: this is a trivial but useful observation to comment Giunta's inferences.  On account of the rather uncontrollable behavior of sulfur, which gives it a predominant weight in the total deviation from the average,  I did not regard as appropriate to include its $C_p$ value in the statistical analysis that follows\footnote{Giunta rises a question of allotropy for tin too, which turns from the usual ductile `white' phase to the brittle `grey' tin by lowering the temperature below $13.2^\circ$. However, this is true only for very pure tin, since even a small amount of impurity lowers a lot the transition temperature. Besides, the transition kinetics is very slow even for pure tin (namely, white tin is highly metastable). Therefore such a phase change, known in France as \emph{la l\`{e}pre d'\'{e}tain}, may well have ruined the tin buttons of the uniforms of Napoleon soldiers during the bitter winter of the long Russian campaign, as recently suggested, but it would have hardly took place within the limited time of a cooling experiment. Hence, very likely Petit and Dulong measured \emph{white} tin.}.

Compare first the means of the three data set, which show that the values obtained by Regnault are on the average 6\% higher than those by Petit and Dulong. Regnault himself pointed out this difference, arguing that some details of the experimental protocol used by Petit and Dulong, in particular for those elements that they also measured with the method of mixtures, may have lead them to underestimate the heat capacity\footnote{\label{Reg}Regnault, however, had no doubt on the fact that Petit and Dulong \emph{did} perform the measurements presented in 1819 with the method of cooling. He also points out one of the main cause of errors of this technique, namely the possible condensation or vapor on the blackened inside of the vessel, which would reduce its absorbance. His attentive description suggests that he may well have \emph{seen} the apparatus shown in Fig. 4, arguably when he was a student at the \'{E}cole, directed at that time by Dulong.}. Unfortunately, he does not seem to be right. While the values he obtained are on the average higher than the modern ones by 5\%, this figure decrease to a skimpy 0.7\% (in the opposite direction) for the Petit and Dulong's data.

But what Giunta mostly cares about are the fluctuations about the mean. Admittedly, the situation here is suspicious, since the relative standard deviation (the coefficient of variation) $CV =s/\langle c\rangle$ of the data of Petit and Dulong ($CV_1 = 0.013$) is about twice larger than that of modern data ($CV_0 = 0.027$)\footnote{If we include sulfur, this figure rises to about 3.}. To an experimentalist like me, this surely smells of data adjustment, although the fact that the relative standard deviation of Regnault's data ($CV_2 = 0.032$) is only marginally higher than the modern one would also imply that his experimental precision was comparable with that reached 80 years later, which is also a bit strange\footnote{Including sulfur, Regnault's data dispersion becomes even \emph{smaller} than the modern one.}. While data adjustment is a reasonable hypothesis, data fabrication, however, is much less credible: Improving data precision by adjusting them is one thing, another one is increasing their \emph{accuracy}, if you do not have a benchmark reference. Namely, how could Petit and Dulong `divine' their data so well as to agree much better than Regnault with a modern data set that was long to come?

To support his claim that the Dulong and Petit data have not been adjusted but truly `fabricated', Giunta compares their variance $\sigma^2$ with that one  of modern values by means a standard $F$-test, which seems to support a null hypothesis, namely that there is no relationship between the two sets of data. However Snedecors's $F$-test applies only to populations that are \emph{normally} distributed. As I warned before, this is far from being the case\footnote{Giunta claims to have \emph{tested} that the two data sets are normally distributed. I guess none of my colleagues would be so daring with only 13 data points.}. At the cost of sounding pedantic, let me stress again that the differences in the observed values of $c_p$ at a given temperature are not due to `errors': their are physical \emph{facts}, they cannot be reduced by improving measurements! Hence, using an $F$-test (but even more refined statistical approaches) to test a null hypothesis makes little sense.

\begin{figure}
\includegraphics[width=\columnwidth]{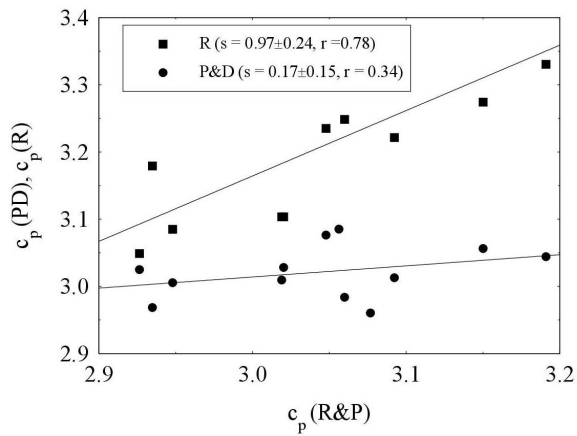}
\caption{\small Molar specific heats obtained by Petit and Dulong (bullets) and by Regnault (squares), pltted versus the values given by Rolla and Piccardi~\cite{Rolla1929}. The slopes $s$ and correlation coefficients $r$ of the linear fits to the two sets of data are shown in the legend.}
\label{f7}       			
\end{figure}
We can nevertheless test \emph{correlations} between the two sets of data, which Fig.~(\ref{f6}) seems to suggest. The extent to which the molar specific heats by Petit and Dulong are linearly correlated with those in Ref.~\cite{Rolla1929} can be estimated from Fig.~\ref{f7}, where Regnault data are also plotted for comparison. While the slope $s$ of the linear fit to the P\&D data basically vanishes within the error bar for $s$ (which of course means that the average of the P\&D data is pretty close to the modern one), the value of the correlation coefficient $r\simeq 0.34$ witness a moderate degree of correlation. Testing the significance of $r$ for a set of $n\times n$ data pints \{x,y\} is usually done by transforming  to the variable $t = [(n-2)r^2/(1-r^2)]^{1/2}$ that, \emph{provided that $x$ and $y$ have a bivariate normal distribution}, has a Student's $t$-distribution with $n-2$ degrees of freedom. It seems therefore that here we are incurring in the same problem (non-Gaussianity) we pointed out before. However, in this case stating approximate confidence levels is relatively safer. Baudinet--Robinet has indeed shown  that for a sample of uncorrelated data of size $n \ge 10$, $t$ has approximately the same Student's distribution, \emph{regardless} of the parent distributions of $x$ and $y$~\cite{Baudinet1974} A simple application to our case ($n=12$, $t\simeq 1.146$) shows that, in spite the relatively small value of $r$, the null hypothesis (namely, that the two set of values are fully uncorrelated) can still be rejected with a level of confidence larger than 70\%. Which is not very high, but far from being negligible. Paying Regnault his due, it is however important to notice that the correlation of \emph{ his own } data with the modern ones ($r \simeq 0.78$) is extremely high, which definitely vouches for his scientific fairness.

That Petit and Dulong adjusted their data, or at least selected among those they obtained the `best' ones, possibly fearing that someone could have `stolen' their result\footnote{That Dulong and Petit feared plagiarism quite a lot is evident from a letter of Berzelius to Alexandre Marcet, dated 27 April 1819 (just 8 days after Dulong's presentation at the Academy), where he writes \emph{``Although I am very close to Mr. Dulong, I did not want to get an in--depth knowledge of his work, the details of which have not yet been communicated (the memory read at the Academy was only a preview for deterring the thieves whom Paris is supposed to be full of and to preserve the priority of the discovery), I avoided it because, being myself engaged in the publication of a little work on corpuscular theory, it might well be suspected to have taken advantage of the advice given by Mr. Dulong, and although the contrary is not difficult to prove, since my memory has already been published a year ago in Swedish, I prefer not to be put in the condition of raising any suspicions}"~\cite{Soderbaum1915}.}, is nevertheless a concrete possibility, which would also explain why a careful experimentalist like Dulong was, according to Dumas, so reluctant at making the big step. Conversely, that they cheat to the point of making up 8 results over 13 sounds speculative, not supported by any serious proof, and definitely incongruous with the point of view of valuable contemporary witnesses.
Hence, if I were a judge called upon to decide whether on 19 April 1819 Petit and Dulong perpetrated a gross `scientific fraud' by fabricating most of their data, I would surely set them free  for insufficient evidence, although I might not, in good conscience, fully acquit them for having not committed the crime. The trial, to me, is over.

\section{Legacy}
The fate of the DP `law' and its effects on the development of physics are well known. By the middle of the XIX century Regnault had already shown that elements with a low atomic weight and high melting temperatures like boron, carbon, and silicon had exceptionally low specific heats at room temperature. Yet, in 1875 Heinrich Weber showed that even the heat capacity of these elements approaches at high temperature the value predicted by the DP law, which should then be regarded as a \emph{limiting} law,\footnote{Weber's observation was crucial for Thomas Humpidge to find in 1885 the correct atomic weight of Beryllium (at that time also called `glucynum'), which shows a very anomalous value of $c_p$ too\cite{Humpidge1885}.}. In his words~\cite{Weber1875},
\begin{quotation}
\noindent \emph{The three curious exceptions to the Dulong-Petit law which were until now a cause for despair have
been eliminated: the Dulong-Petit law for the specific heats of solid elements has become an unexceptional rigorous law.}
\end{quotation}
Yet, it became very soon clear that Weber's attempt to `rescue' the DP law was just a way of getting round the real problem. After all his own data, obtained by cooling with dry ice, showed that the specific heat of diamond went down by more than one order of magnitude by decreasing $T$ from $1000$ to $-50^\circ$C. In 1905, when Dewar managed to reach temperatures as low as 20\,K using liquid hydrogen, it became evident that
the specific heat actually vanishes as $T\rightarrow 0$, which of course was totally incompatible with classical statistical mechanics\footnote{Although a \emph{moderate} temperature dependence could be justified because of anharmonic effects, as already pointed out by Richarz~\cite{Richarz1885}.} The vanishing of $c_p$ low temperature behavior predicted by the Einstein model, a triumph of the early quantum physics, the introduction of collective vibrations by Debye that provided the correct $T^3£$ limiting behavior, the subsequent refinement by Born and van K\'{a}rm\'{a}n  that paved the way to the modern investigation of phonons in solids, is a known story to physicists (for an accurate review, see~\cite{Blackman1941}.

Our community is probably less acquainted with the at least equally important role played by the DP law in chemistry, in particular as a key tool to unravel the nature of the atomic weights. We have already seen how Berzelius used the law to correct some, but not all, of the atomic weights he had measured. What Berzelius could not accept at all, however, was the law proposed in 1811 by Amedeo Avogadro~\cite{Avogadro1811}, stating that a given volume of any gases, for fixed values of temperature and pressure, always contained the same number of molecules.  When put together with the gas atomic weights obtained by Dumas (see footnote~\ref{Dumas}), Avogadro's law implied indeed that even simple gases like hydrogen or nitrogen had to be made of \emph{diatomic} molecules. For Berzelius, who believed that bonds between atoms always derive from electric forces, this was clearly untenable and almost preposterous: how could two identical atoms with the same charge bind? As a matter of fact, Avogadro's ideas remained in
oblivion for a long time\footnote{Actually, the nomenclature used by Avogadro, who always refused to use the word `atom', did not help. For instance, what we now call an atom was an `elementary molecule', while `constituent'  and `integral' molecules were respectively the molecule of a pure element and of a compound of different atoms.} until they were given the place they deserve by Stanislao Cannizzaro, the greatest Italian chemist of the XIX century who had studied calorimetry with Regnault at the Coll\`{e}ge de France~\footnote{Cannizzaro went to France to escape from a death penalty he had been sentenced to for having actively participated to the 1848 Sicilian revolution against the Bourbon rulers.}.  In a letter to the secretary of \emph{Il Nuovo Cimento } Salvatore De Luca~\cite{Cannizzaro1858}, entitled \emph{Sunto di un corso di Filosofia Chimica}\footnote{Cannizzaro's seminal letter is meant to be just the summary of the chemistry course he taught  at the University of Genova.}, Cannizzaro, granting Avogadro's hypothesis, made \emph{extensive} use of the DP law to evaluate accurate atomic and molecular weights. This led him to formulate his fundamental result, a major step towards giving physical reality to atoms:
\begin{quotation}
\noindent \emph{The various quantities of the same element contained in different molecules are all multiples of the same quantity that, always entering as a whole, must be called atom.}\footnote{``Le varie quantit\`{a} dello stesso elemento contenute in diverse molecole son tutte multiple intere di una medesima quantit\`{a} la quale, entrando sempre intera, deve a ragione chiamarsi atomo".}
\end{quotation}
 Cannizzaro presented his ideas at the first global conference of chemists, held in Karlsruhe in 1860. One of the conference's participants was a young Russian chemistry student, Dmitrii Ivanovich Mendeleev, who, along with everyone else in attendance, received a copy of the paper by Cannizzaro. Immediately after reading the paper, Mendeleev wrote an enthusiastic letter to his teacher A. A. Voskresenskii in St. Petersburg, mentioning him that he found that all Cannizzaro's values satisfied the DP law~\cite{Kaji2002}. The appreciation of Cannizzaro's work and of the importance of the DP law had a crucial consequence in the construction of the periodic table. Mendeleev indeed used the DP law to correct the atomic weights for indium, cerium, and uranium, which were wrong in his first 1869  table\footnote{The 1869 values, $m_a(\mathrm{In}) = 75.6$, $m_a(\mathrm{Ce}) = 92$, $m_a(\mathrm{U}) = 116$ were changed to $m_a(\mathrm{In}) = 113$, $m_a(\mathrm{Ce}) = 138$, $m_a(\mathrm{U}) = 240$, which, albeit not perfect, allowed him to position these elements in the correct groups of the periodic table.} allowing him to  produce the remarkable table of 1871 that lasted basically unchanged for almost 70 years~\cite{Kaji2002,Brooks2002}.

 As I already stated, I am strongly convinced that they can be charged at most of misdemeanour rather than of a scientific `crime'. But even if I am totally wrong, even if they really deceived, this has been one of the most fruitful frauds of the history of science, because the legacy of the work jointly performed by an exquisite experimentalist like Dulong and a gifted theorist like Petit is truly invaluable.

 \section*{Acknowledgements} This investigation would not have been possible without the precious support of free online repositories like Gallica, the Internet Archive, and the Hathi Trust Digital Library, which provide an invaluable service not only to the community of professional scholars, but also to anyone like me who is just curious of the way our science blossomed.

\end{document}